\documentclass[aoas,MSNbibl,nameyear,dvips]{arximspdf}
\usepackage{graphicx}

\doi{10.1214/12-AOAS546} 
\volume{6}
\issue{3}
\pubyear{2012}
\firstpage{1256}
\lastpage{1279}

\makeatletter

\newtheorem{Proposition}{Proposition}
\newtheorem{Lemma}{Lemma}
\newtheorem{Theorem}{Theorem}
\newtheorem{Claim}{Claim}

\newproclaim{Assumption}{Assumption}

\makeatother

\begin{document}
\begin{frontmatter}

\title{Order selection in nonlinear time series models
with application to the study of cell memory\thanksref{T1}}
\runtitle{Nonlinear time series modeling}

\thankstext{T1}{Supported by NSF Grants DMS-09-05753 and CMMI 0927572.}

\begin{aug}
\author[A]{\fnms{Ying} \snm{Hung}\corref{}\ead[label=e1]{yhung@stat.rutgers.edu}}
\runauthor{Y. Hung}
\affiliation{Rutgers University}
\address[A]{Department of Statistics\\
\quad and Biostatistics\\
Rutgers University\\
Piscataway, New Jersey 08854\\
USA\\
\printead{e1}} 
\end{aug}

\received{\smonth{7} \syear{2011}}
\revised{\smonth{1} \syear{2012}}

%
\begin{abstract}
Cell adhesion experiments are biomechanical experiments studying the
binding of a cell to another cell at the level of single molecules.
Such a study plays an important role in tumor metastasis in cancer
study. Motivated by analyzing a repeated cell adhesion experiment, a
new class of nonlinear time series models with an order selection
procedure is developed in this paper. Due to the nonlinearity, there
are two types of overfitting. Therefore, a double penalized approach is
introduced for order selection. To implement this approach, a global
optimization algorithm using mixed integer programming is discussed.
The procedure is shown to be asymptotically consistent in estimating
both the order and parameters of the proposed model. Simulations show
that the new order selection approach outperforms standard methods. The
finite-sample performance of the estimator is also examined via a
simulation study. The application of the proposed methodology to a
T-cell experiment provides a better understanding of the kinetics and
mechanics of cell adhesion, including quantifying the memory effect on
a repeated unbinding force experiment and identifying the order of the
memory.
\end{abstract}

%
\begin{keyword}
\kwd{Consistency}
\kwd{micropipette experiment}
\kwd{order selection}
\kwd{single molecule}
\kwd{threshold autoregressive model}.
\end{keyword}

\end{frontmatter}

\section{Introduction}\label{sec1}

Cell adhesion plays an important role in many physiological and
pathological processes, especially in tumor metastasis in cancer study.
Cell adhesion experiments refer to biomechanical experiments that study
the binding of cells at the molecular level. The binding is mediated by
specific interaction between cell adhesion proteins, called receptors,
and the molecules that they bind to, called ligands. The resulting bond
is called the receptor-ligand bond. There are various types of
measurements in the cell adhesion experiments to study different
aspects of the binding, such as the binding frequency and bond lifetime
measurements [\citet{Zaretal07}, \citet{Huaetal10}]. This
research is inspired by analyzing a specific type of cell adhesion
experiment known as the unbinding force assay [Marshall et~al.
(\citeyear{Maretal03}, \citeyear{Maretal05})].

\begin{figure}

\includegraphics{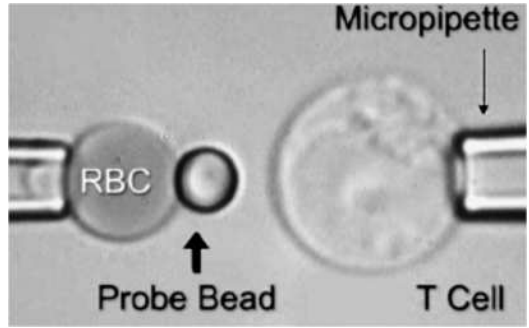}

\caption{Illustration of the biomembrane force probe.}\label{BFP}
\end{figure}

\begin{figure}[b]

\includegraphics{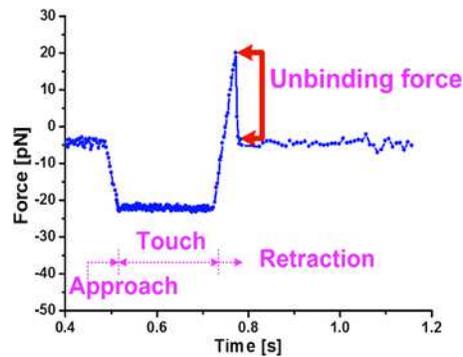}

\caption{One cycle of the unbinding force experiment.}\label{Onecycle}
\end{figure}

Receptor-ligand bonds that mediate
cell adhesion are often subjected to forces that regulate their
dissociation; therefore, an important issue is to study the unbinding
force of
a receptor-ligand bond.
To address this issue, the unbinding force assay is developed by using
a high-tech version of the micropipette known as the
biomembrane force probe [\citet{Cheetal08}]. A~biomembrane force
probe is illustrated in Figure~\ref{BFP} where a probe bead (left) is
attached to the apex of the micropipette-aspirated red blood cell to
allow tracking of the deflection of another cell (right).
Figure~\ref{Onecycle} illustrates one cycle of the unbinding force
assay. It includes an approaching stage where the probe bead and the
T-cell are brought into contact. In the next stage, the touch of the
two subjects is controlled with a~given contact time so that
a~receptor-ligand bond might occur. In the last stage, the probe bead and
the T-cell are retracted at a constant rate until they go back to the
unbinding position that indicates the bond failure. The $y$-axis in
Figure~\ref{Onecycle} represents the applied force in the foregoing
process. The unbinding force is measured by the force difference
observed at the point of bond failure.

Two interesting questions are raised in analyzing the repeated
unbinding force tests where the unbinding force assay (i.e.,
approaching, contact and retraction) is performed repeatedly for each
pair of experimental units, including a T-cell and a probe bead
attached to a red blood cell.
Such repeated assays are conducted for different pairs of units as
replicates. The objective of the experiments is to study the dependence
of the repeated unbinding force measurements because it was discovered recently
that cells appear to have the ability to ``remember'' the previous
adhesion events. \citet{Zaretal07}, \citet{Hunetal08} and
\citet{Huaetal10} demonstrated that in some biological systems the
occurrence of binding in the immediate past assay could either increase
or decrease the likelihood for the next assay to result in a binding.
Such memory effects can affect not
only through the binding frequency but also the unbinding force.
Hence, the first question is how to model the memory effect on the
repeated unbinding force assays.
Apart from this, different receptor-ligand bonds can have a different
order of the memory due to their string strength difference. Specifying
the order of the memory for receptor-ligand bonds is important because
it can be used to classify the bonds into groups for further biological study.
Therefore, the other question is how to identify the order of the memory.

\begin{figure}

\includegraphics{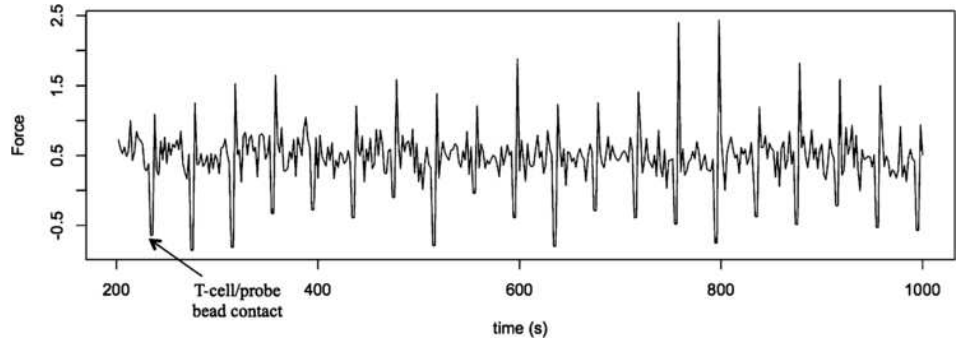}

\caption{The measurements from the repeated unbinding force
assays.}\label{Adhesion}
\end{figure}

To answer the foregoing questions, a naive approach is to study the
memory on the unbinding force by a time series model. However, the
standard time series models cannot be applied directly. The reason is
as follows. Due to the inherent stochastic nature of single molecular
interaction, any particular assay has two random outcomes, either a
receptor-ligand bond occurs or not. An unbinding force is
representative and the resulting memory effects are considered only if
the corresponding assay is associated with the occurrence of a bond.
Theoretically, a distribution function might be used to capture the
chance of a bond formation with respect to unbinding force. However,
the related studies are mainly developed based on the independent
assumption on the repeated adhesion experiments
[\citet{Maretal05}]. Being the first attempt to study the memory, we assume that
the occurrence of a bond is determined by having the unbinding force
above some threshold, which can be interpreted as the average unbinding
force for bond dissociation. That is, if a bond occurs during the
contact, the unbinding force would be larger than some threshold. The
threshold, however, is unknown and has to be estimated from the data
because of the detection limits and measurement errors.
For example, Figure~\ref{Adhesion} is an example of the experiments
%
with 20 repeated unbinding force assays generated from
\citet{Hunetal08}. For each cycle of the assay, unbinding forces can be easily
measured as described in Figure~\ref{Onecycle}. A threshold has to be
determined so that time series
models can be applied\vadjust{\goodbreak} to those forces that are above the threshold.
Failing to include such a threshold term can lead to a systematic bias
in the successive adhesion assays. Because of the unknown threshold,
conventional time series modeling techniques cannot be used.
Furthermore, to identify the order of the memory, a new order selection
approach that takes into account the foregoing features is called for.

A new time series model is proposed in this article to study the memory
effect on the repeated unbinding force assays. It is a multiple
nonlinear time series model with an unknown threshold parameter. Even
though there are numerous studies on nonlinear time series modeling
[\citet{TonLim80}, \citet{Tsa89}, \citet{FanYao03}],
most of them are developed based on a~single series of observations and
focus on the situation where nonlinearity is determined by a particular
variable. For example, the threshold autoregressive model [Tong
(\citeyear{Ton83}, \citeyear{Ton07})] is constructed for a single
series of observations with a delay parameter indicating the variable
where the threshold is applied. The proposed nonlinear model is
different from the existing nonlinear time series models in that there
is no specific delay parameter involved. Instead, the threshold is
applied to all the historical observations. Moreover, there is a
hierarchical structure imposed upon the nonlinear model that makes the
model more interpretable. Besides, this model handles multiple time
series by incorporating random effects to take into account the
heterogeneity among experimental units.

Identifying the order of the memory is equivalent to specifying the
correct order of the proposed time series model. This is different from
standard order selection problems because there are two types of
overfitting associated with the proposed nonlinear time series model.
Thus, a double penalized approach is developed and a global
optimization algorithm using mixed integer programming (MIP) is
introduced to implement this approach. The order selection consistency
and asymptotic properties for the proposed method are discussed. The
discontinuity of the conditional mean function of the new model results
in nonstandard asymptotics for the estimators.\vadjust{\goodbreak}

Although the methodology is motivated by the analysis of biomechanical
experiments, it can be applied to a wide variety of studies, such as
longitudinal data analysis [\citet{Digetal02}], econometrics and
influenza modeling. For example, in influenza modeling
[\citet{HymLaF03}], the proposed method can be applied to model
the spread of a disease, such as SARS. Because an epidemic threshold is
used to indicate the take off and die out of an epidemic, the spread of
the disease is of interest only when the threshold is reached, such as
the infected population exceeding some amount. These thresholds are
often unknown and estimated from the data. Therefore, the proposed
model can be desirable for these studies.

The remainder of the paper is organized as follows. In Section~\ref{sec2} the
nonlinear time series model is introduced. The estimation and order
selection procedures with a global optimization algorithm are
introduced. In Section~\ref{sec3} the order selection consistency and some
asymptotic properties of this model are discussed. The performance of
the new model and the order selection procedure is demonstrated via
simulations in Section~\ref{sec4}. The proposed model is applied to an unbinding
force assay in Section~\ref{sec5}. Summary and concluding remarks are given in
Section~\ref{sec6}.

\section{New class of nonlinear time series models}\label{sec2}

\subsection{Modeling}\label{sec2.1}

A new multiple nonlinear time series model is introduced in this section.
Assume $y_{it}$ represents the unbinding force observed from the $i$th
subject at time $t$, where $i=1,\ldots,n$,
$t=1,\ldots,m$ and the sample size $N=mn$. Define $\tau$ as a threshold
parameter. Having the unbinding force above $\tau$ indicates that the
corresponding contact results in a receptor-ligand bond and no bond otherwise.
A random effect $\bolds{\alpha}=(\alpha_1,\ldots,\alpha_n)$ is
incorporated to take into account a variety of situations with the
multiple time series, including subject
heterogeneity, unobserved covariates and other forms of overdispersion.
The random effects $\alpha_i$'s are assumed to be mutually independent
and normal distributed with mean ${0}$ and variance $\sigma^2$ in this
paper. The following model is proposed to quantify the memory effect on
the unbinding forces that are associated with receptor-ligand bonds:
%
\begin{equation}\label{NLTeq1}\qquad
\cases{
y_{it}=\alpha_i+\beta_0+\varepsilon_{it},&\quad if
$y_{i,t-1}\leq\tau$,\vspace*{1pt}\cr
y_{it}=\alpha_i+\beta_0+\beta_1 y_{i,t-1}+\varepsilon_{it},&\quad if
$y_{i,t-1}> \tau, y_{i, t-2}\leq\tau$,\vspace*{1pt}\cr
y_{it}=\alpha_i+\beta_0+\beta_1 y_{i,t-1}\vspace*{1pt}\cr
\hphantom{y_{it}=}{}+\beta_2 y_{i,t-2}+\varepsilon
_{it}, &\quad if $y_{i,t-1}> \tau, y_{i, t-2}> \tau,
y_{i,t-3}\leq\tau$,\vspace*{1pt}\cr
\hphantom{y_{it}}\hspace*{5.5pt}\vdots&\quad\vdots\cr
y_{it}=\alpha_i+\beta_0+\beta_1 y_{i,t-1}\vspace*{1pt}\cr
\hphantom{y_{it}=}{}+\cdots+\beta_k
y_{i,t-k}+\varepsilon_{it}, &\quad if $y_{i,t-1}> \tau,\ldots, y_{i,
t-k}> \tau$,}
\end{equation}
where $\beta_i$'s are the fixed effects and the error terms $\varepsilon
_{it}$ are independent with distribution $N(0,\sigma^2_{\varepsilon})$.

The first equation in~(\ref{NLTeq1}) corresponds to the situation where
no receptor-ligand bond occurs in the previous test (i.e., $y_{i,t-1}
\leq\tau$). It amounts to modeling the unbinding forces in a sequence
of independent adhesion tests. Let the mean unbinding force be $\beta
_0$. The estimated value for $\beta_0$ is the average unbinding force
in independent adhesion assays and can change with different settings
of the experimental variables, such as different contact durations.
Extensions can be easily achieved by incorporating these experimental
variables into the model. The second equation in~(\ref{NLTeq1})
describes the unbinding force when a receptor-ligand bond occurs in the
previous test (i.e., $y_{i,t-1}>\tau$) but no bond in $y_{i,t-2}$
(i.e., $y_{i,t-2} \leq\tau$). In this situation, a memory could be
carried over from the previous observations. Thus, a~first-order
autoregressive model is considered. This autoregressive modeling
continues to the previous $k$ assays. Similar
interpretation can be given to the rest of the model. The value $k$
represents the upper bound of the memory order; detailed discussions on
identifying the order of the memory are given in Section~\ref{sec2.2}.

The above model can be written in a concise form as follows:
%
\begin{eqnarray}\label{MUTAREmain}\quad
y_{it}&=&{\mathbf z}'_{i} \bolds{\alpha}+\beta_0+\beta_1 y_{i, t-1} {
I}[y_{i,t-1}>\tau]+\beta_2 y_{i, t-2} {I}[y_{i,t-1}>\tau, y_{i,t-2}>
\tau]\nonumber\\
&&{}+\cdots+\beta_k y_{i, t-k} {I}[y_{i,t-1}>\tau, \ldots, y_{i,t-k}>
\tau
]+\varepsilon_{it}\\
&=&g(\bolds{\beta},\tau,\sigma^2 \mid
H_{it})+\varepsilon_{it},\nonumber
\end{eqnarray}
where $I(y_{i,t-1}>\tau)$ is an indicator function which takes value
one if $y_{i,t-1}>\tau$ and zero otherwise. The fixed effects are
denoted by $\bolds{\beta}=(\beta_0,\beta_1,\ldots,\beta_k)'$, the
information from previous observations are included in $H_{it}=(1,
y_{i,t-1},\ldots,\break y_{i,t-k})$, and ${\mathbf z}_i=\{z_{i,1},\ldots,z_{i,n}\}
'$ is the design matrix for the random effects $\bolds{\alpha}$
such that ${\mathbf z}'_{i} \bolds{\alpha}=\alpha_i$. Since the
proposed model is not limited to the analysis of unbinding force assay,
the random intercept alone may not be sufficient to capture the
variation exhibited in other applications. Hence, we use a~general
random effect structure hereafter. We call this new nonlinear time
series model the multiple threshold autoregressive (MUTARE) model.

The MUTARE model is very general and includes an interesting special
case with a single series of observations. Assuming that the time
series observations are~$y_t$, $t=1,\ldots,m$, the special case of the
MUTARE model can be written as
%
\begin{eqnarray}\label{NLTnoRE}
y_{t}&=&\beta_0+\beta_1 y_{t-1} { I}[y_{t-1}>\tau]+\cdots\nonumber\\[-8pt]\\[-8pt]
&&{}+\beta_k y_{t-k}
{I}[y_{i,t-1}>\tau, \ldots, y_{t-k}> \tau]+\varepsilon_{t}.\nonumber
\end{eqnarray}
This is different from the conventional nonlinear time series models.
The closest model in the literature is the threshold autoregressive
models introduced
by Tong (\citeyear{Ton83}, \citeyear{Ton90}). There are various extensions of the threshold
autoregressive models [\citet{SamChaSte07}] and the nonlinearity
therein is determined by a particular variable with which the threshold
parameter is defined.\vadjust{\goodbreak}
The \mbox{MUTARE} model, however, has the threshold applied to all the
historical observations. Furthermore, different from the threshold
autoregressive model where piecewise linear submodels are fitted
separately, a hierarchical structure is imposed upon the submodels in
MUTARE as illustrated in~(\ref{NLTeq1}), which makes the model easier
to interpret.\vspace*{-1pt}

\subsection{Estimation and order selection procedure}\label{sec2.2}

A crucial step in this study is to specify the order of the memory,
denoted by $k_0$. This is an order selection problem but different from
standard ones in that there are two types of overfitting. By maximizing
the log likelihood function, the resulting model may overfit the data
with some small values of nonzero $\beta_j$'s (type I overfitting)
and/or with a large estimated order (type II overfitting). This is not
surprising given the same problem experienced in estimating parameters
in finite mixture models [\citet{CheKha08}]. Therefore, we
propose to penalize type I overfitting by a function $ P_{\lambda
_1}(|\beta_k|)$ and penalize type II overfitting by the estimated order
($\max_j \{j\dvtx \beta_j \neq0\}$). The reason to consider type II
overfitting is because the MUTARE model has a hierarchical structure as
shown in~(\ref{NLTeq1}). Once the order of the model (i.e.,
$\max_j \{j\dvtx \beta_j \neq0\}$) is determined, all the previous
equations have to be considered. So a double penalized likelihood is
defined as
%
\begin{equation}\label{PenMLE}
\operatorname{pl}(\bolds{\beta}, \sigma^2, \tau)=2 \log L(\bolds{\beta},
\sigma^2, \tau)-\sum_{j=1}^k P_{\lambda_1}(|\beta_j|)-\lambda_2 \max_j
\{j\dvtx \beta_j \neq0\},
\end{equation}
where $L$ is the likelihood function. By maximizing~(\ref{PenMLE}), the
solutions, $\hat{\bolds{\beta}}$ and $\max_j \{j\dvtx \hat{\beta}_j
\neq0\}$, are the estimated parameters and order of the memory.

To prevent the first type of overfitting, there are different penalty
functions discussed in the literature [\citet{DonJoh94},
Tibshirani (\citeyear{Tib96}, \citeyear{Tib97}), \citet{FanLi01}]. Here we focus on the
adaptive Lasso [\citet{Zou06}] where
$P_{\lambda_1}(|\beta_j|)=\lambda_1 \nu_j |\beta_j|$ and $\nu_1,\ldots
,\nu_k$ are known weights. The specification of $\nu_j$ can be fairly
flexible and more discussions can be found in \citet{Zou06}. We consider a
weight vector suggested in \citet{Zou06} with $\hat{\nu}_j=|\hat{\beta
}_j|^{-\rho}$, where $\rho>0$ and $\hat{\beta}_j$ is a
root-$n$-consistent estimator of $\beta_j$. In \citet{Hun}, it is shown
that the MLE of $\bolds{\beta}$ is root-$n$-consistent under model
(\ref{MUTAREmain}), therefore it can be applied.

By the following proposition, we can have a closer look at how the
double penalized approach works. The proof is straightforward and is omitted.\vspace*{-1pt}
\begin{Proposition}\label{Proposition1}
The penalized likelihood
function in~(\ref{PenMLE}) is equivalent~to
%
\begin{eqnarray}\label{PenMLE3}\qquad
\operatorname{pl}(\bolds{\beta}, \sigma^2, \tau)&=&2 \log L(\bolds{\beta},
\sigma^2, \tau)-\sum_{j=1}^k P_{\lambda_1}(|\beta_j|)-\lambda_2 \sum
_{j=1}^k I(\beta_j \neq0)\nonumber\\[-8pt]\\[-8pt]
&&{}-\lambda_2 \sum_{j=1}^k I(\beta_j =0\mbox{, at least
one }\beta_{j+p}\neq0, p=1,\ldots,k-j).
\nonumber\vadjust{\goodbreak}
\end{eqnarray}
\end{Proposition}

Equation~(\ref{PenMLE3}) connects the penalty for type II overfitting
with the $L_0$ penalty, which directly controls the number of nonzero
coefficients in the model. Therefore, the double penalized approach is
closely related to a combination of $L_0$ and $L_1$ penalties, which is
carefully studied by \citet{LiuWu07} and found to deliver better
variable selection than the $L_1$ penalty while yielding a more stable
model than the $L_0$ penalty.

\subsection{Mixed integer programming}\label{sec2.3}

In this section a global optimization algorithm is introduced using the
idea of MIP. MIP is an active research area in operations research with
many applications. The objective here is to solve $\beta_j$'s by
maximizing the double penalized likelihood function~(\ref{PenMLE}). It
is achieved by the following proposition.
\begin{Proposition}\label{Proposition2}
The penalized likelihood
function in~(\ref{PenMLE}) is equivalent~to
%
\begin{eqnarray}\label{PenMLE2}
\operatorname{pl}(\bolds{\beta}, \sigma^2, \tau)&=&2 \log L(\bolds{\beta},
\sigma^2, \tau)-\sum_{j=1}^k P_{\lambda_1}(|\beta_j|)\nonumber\\[-8pt]\\[-8pt]
&&{}-\lambda_2 \sum
_{j=1}^k \bigl(1-I(\beta_j=\cdots=\beta_k=0)\bigr).\nonumber
\end{eqnarray}
\end{Proposition}

As discussed in Proposition~\ref{Proposition2}, this problem is equivalent to the
maximization of~(\ref{PenMLE2}). Substitute variable $\beta_j$ by two
nonnegative variables $\beta_j^+$ and $\beta_j^-$ with $\beta_j=\beta
_j^+-\beta_j^-$. Then, we have $|\beta_j|=\beta_j^++\beta_j^-$, and the
maximization problem in~(\ref{PenMLE2}) can be converted into a MIP
problem with maximization of
\[
2 \log L(\bolds{\beta}^+-\bolds{\beta}^-,\sigma^2, \tau)-\sum
_{j=1}^k P_{\lambda_1}(\beta_j^++\beta_j^-)-\lambda_2 \sum_{j=1}^k z_j,
\]
subject to
\begin{eqnarray*}
\beta_1^++\beta_1^-+\beta_2^++\beta_2^-+\cdots+\beta_k^++\beta_k^-
&\leq& Mz_1,\\
\beta_2^++\beta_2^-+\cdots+\beta_k^++\beta_k^-
&\leq& Mz_2,\\
&\vdots&\\
\beta_k^++\beta_k^- & \leq & Mz_k, \\
\beta_j^+,\beta_j^- & \geq & 0,\qquad j=1,\ldots,k,\\
z_j &\in& \{0,1\},
\end{eqnarray*}
where $M$ is a very large constant and we can choose it to be the
smallest upper bound of $\sum_j |\beta_j|$ if the prior knowledge is
available. In the simulations, we apply the setting $M=50$ and it works
reasonably well in practice. In general,\vadjust{\goodbreak} $M$ can be even larger (e.g.,
$M=1000$) for those problems with large $k$. Note that since $\beta
_j^++\beta_j^-$ are to be minimized, $\beta_j^+$ and $\beta_j^-$ would
not be both positive in the optimal solution.

To solve the foregoing MIP problem, there are numerous methods such as
the most popular branch-and-bound algorithm. More details about
algorithms and the related issues can be found in
\citet{NemWol99}. The examples we considered in this article are
solved by the C language with a GLPK package (available at
\url{http://www.gnu.org/software/glpk}). Some other commercial
optimization software such as CPLEX is also available to solve such a
problem. The complexity of MIP can be considerably affected by
introducing too many integer variables (i.e., $z_j$'s), but it is in
general not a critical concern. This is because the number of integer
variables incorporated increases with the order $k$, and it is usually
in a manageable size in this application. For other applications with a
large value of $k$, one can obtain a reasonably good solution (not
necessarily optimal) by setting a restriction on the computing time to
achieve efficiency.

Next we discuss the choice of the tuning parameters, $\lambda_1$,
$\lambda_2$ and $\rho$. There are different approaches available in the
literature for selecting tuning parameters [\citet{Sto74},
\citet{CraWah78}, \citet{FanGij96}]. \citet{BurChoNol94}
introduced the $h$-block cross-validation for dependent data. The idea
is to modify the leave-one-out cross-validation and reduce the training
set by removing the $h$ observations preceding and following the
observation in each test set. Such blocking allows near independence
between the training and test set. This approach is further improved by
\citet{Rac00} to achieve asymptotic consistency. That is, instead
of leave-one-out, the size of the validation set is increased to $n_v$.
So the training set has size $n_c$ and $n_v+n_c+2h=m-k$. In this paper,
we implement Racine's approach with the setting $h=(m-k)/4$ and $n_c$
being the integer part of~$m^{0.5}$, which appears to work well in a
wide range of situations in practice [\citet{Rac00}].

The rest of the parameters can be estimated by the standard maximum
likelihood approach.
Denote the observation by vector $Y=({\mathbf y}_1,\ldots,
{\mathbf y}_n)'$, where the observations for subject $i$ are denoted by
${\mathbf y}_i=(y_{i1},\ldots,y_{im})'$. Given the historical
information $H_{it}$ and the random effects, the associated likelihood
as a function of the fixed effects $\bolds{\beta}$ and the
threshold parameter can be written as
\[
L(\bolds{\beta},\tau\mid\bolds{\alpha})=\prod_{i=1}^n \prod
_{t=1}^m l(y_{it}\mid\bolds{\alpha}, H_{it}),
\]
where $l(\cdot)$ is the likelihood for each observation $y_{it}$ given
$\bolds{\alpha}$ and the corresponding historical information.
Considering the normality of the error $\varepsilon$ and random
effects~$\bolds{\alpha}$, the joint log likelihood can be easily
derived as
%
\begin{eqnarray}\label{likelihood}
&&
2\log L(\bolds{\beta}, \sigma^2, \tau)\nonumber\\[-8pt]\\[-8pt]
&&\qquad=-{\log}|{\mathbf
W}|-\bigl(Y-g(\bolds{\beta}, \tau, \sigma^2 \mid H)\bigr)'{\mathbf
W}^{-1}\bigl(Y-g(\bolds{\beta}, \tau,\sigma^2 \mid H)\bigr),\nonumber
\end{eqnarray}
where $g(\bolds{\beta}, \tau,\sigma^2 \mid H)$ is the mean vector,
$H=(H'_{1},\ldots,H'_n)'$, $H_i=(H'_{i1},\ldots,\allowbreak H'_{im})'$, $Z$ is the
design matrix\vspace*{1pt} for the random effects with rows ${\mathbf z}'_i$, and
${\mathbf W}=\sigma^2_{\varepsilon}{\mathbf I}+\sigma^2 Z Z'$. Note
that $\sigma ^2_{\varepsilon}$ is assumed to be known for notational
convenience. The variance component $\sigma^2$ is estimated by
maximizing the original likelihood throughout the paper and the
estimator can be further improved by the restricted maximum likelihood
[\citet{McCSea01}]. Such a~version of the variance components
developed for the linear mixed model can be easily extended to the
multiple threshold autoregressive model so that the estimated variance
component is invariant to the values of the fixed effects and the
degrees of freedom for the fixed effects can be taken into account
implicitly.

\section{Large sample properties}\label{sec3}

The consistency of the order selection procedure and the asymptotic
properties of the resulting estimators in the MUTARE model are studied
in this section.
The parameter space of $\bolds{\gamma}=(\bolds{\beta},\tau
,\sigma^2)$ is denoted by $\Omega$ and the true parameter is denoted by
$\bolds{\gamma}_0=(\bolds{\beta}_0,\tau_0,\sigma_0^2)$.
Assumptions and proofs are deferred to the \hyperref[app]{Appendix}.

Lemma~\ref{lemma1} shows that the maximum penalized likelihood estimator for the
\mbox{MUTARE} model is stochastically bounded.
\begin{Lemma}\label{lemma1}
Under Assumptions~\ref{AssumptionA1}--\ref{AssumptionA4}, there exists a $\nu>0$
such that, for $m$ and $m$ sufficiently large, the maximum penalized
likelihood estimator of the parameter $\bolds{\gamma}=(\bolds
{\beta}, \tau,\sigma^2)$ lies in a compact space \mbox{$\Omega_1=\{\bolds
{\gamma} \in\Omega\dvtx |\bolds{\gamma}-\bolds{\gamma}_0|\leq\nu
\}$} almost surely.
\end{Lemma}

The convergence rate of the estimated threshold parameter is derived in
Theorem~\ref{theorem1} for the MUTARE model. This result is analogous to
\citet{Cha93} for the least squares estimator of the threshold
autoregressive model. Not surprisingly, the estimated threshold
parameter in the MATARE model has a fast convergence rate [$O(1/N)$]
which is similar to that in the threshold autoregressive model, and the
fast convergence rate is also due to the discontinuity of the
conditional mean function [\citet{Cha93}, \citet{Han00}].
Note that, as a special case, the estimated threshold parameter in
(\ref{NLTnoRE}) obtains a~convergence rate $O(1/m)$.
\begin{Theorem}\label{theorem1}
Under Assumptions~\ref{AssumptionA1}--\ref{AssumptionA4}, the maximum
likelihood estimator of the threshold has the property that $\hat{\tau}
=\tau_0+O_p (1/N)$, based on the MUTARE model.
\end{Theorem}

Define
$\tilde{H}=(\tilde{H}'_1,\ldots,\tilde{H}'_n)$, $\tilde{H}_i=(\tilde
{H}'_{i1},\ldots,\tilde{H}'_{im})$, and
\[
\tilde{H}_{it}=\bigl(1, y_{i,t-1}I(y_{i,t-1}>\tau),\ldots,y_{i,t-k}
I(y_{i,t-1}>\tau,\ldots, y_{i,t-k}>\tau)\bigr).
\]
Let $\bolds{\beta}_0=(\bolds{\beta}'_{(1)},\bolds{\beta
}'_{(2)})'$, where $\bolds{\beta}'_{(1)}$ is a vector with
all the nonzero parameters and the rest of the parameters are denoted
by $\bolds{\beta}'_{(2)}$.
Furthermore, assume\vadjust{\goodbreak} $\frac{\tilde{H}'{\mathbf W}^{-1}\tilde{H}}{N}
\rightarrow\Lambda$, where $ \Lambda$ is positive definite and can be
written as
\[
\Lambda=\left[
\matrix{
\Lambda_{11} & \Lambda_{12}\cr
\Lambda_{21} & \Lambda_{22}}
\right]
\]
according to $\bolds{\beta}'_{(1)}$ and $\bolds{\beta}'_{(2)}$.

In the next theorem, we show that the penalized likelihood estimator of
$\bolds{\beta}$ enjoys the oracle properties [\citet{FanLi01}],
which indicates the consistency in variable selection and the
asymptotic normality. This result also implies the order selection
consistency of the proposed order selection procedure.
\begin{Theorem}\label{theorem2}
Suppose that $\lambda_1/\sqrt{N} \rightarrow
0$ and $\lambda_1N^{(\rho-1)/2}\rightarrow\infty$. Under Assumptions
\ref{AssumptionA1}--\ref{AssumptionA4}, for any $\eta$, $0<\eta<\infty$, the maximum penalized
likelihood estimator of $\bolds{\beta}$ in the MUTARE model
satisfies the following two properties as $n\rightarrow\infty$ and
$m\rightarrow\infty$:
\begin{longlist}[(ii)]
\item[(i)] $\hat{\bolds{\beta}}_{(2)}=0$ with probability 1,
\item[(ii)]
$
\sup_{|\hat{\tau}-\tau_0| \leq\eta/N, |\hat{\sigma}^2-\sigma_0^2|<\eta
/\sqrt{N}} \sqrt{N} (\hat{\bolds{\beta}}_{(1)}-\bolds{\beta
}_{(1)}) \rightarrow^d N(0, \Lambda_{11}^{-1}).
$
\end{longlist}
\end{Theorem}

Apart from the fixed effects, asymptotic distributions of the estimated
variance components deserve more investigation. Numerous works have
appeared in the literature addressing methods of variance component
estimation in linear models and the associated asymptotic properties
[\citet{Jia96}, \citet{McCSea01}]. Strong consistency of the
estimated variance component in nonlinear mixed effect models
[\citet{Nie06}] is expected to be extended to the MUTARE model. A
rigorous theoretical proof along the lines of \citet{Nie06} is not
attempted here, and remains the subject of ongoing theoretical work.
However, it is briefly noted that the asymptotic conditions, such as
Assumptions~\ref{AssumptionA3} and~\ref{AssumptionA4}, required for the results here are indeed met by
the requirement in \citet{Nie06}. The requirement of
$n\rightarrow\infty$ for the main theorems is based upon the asymptotic
study in \citet{Nie06} and it is expected to be further relaxed by
the techniques developed in \citet{Jia96}.

\section{Finite-sample performance and empirical application}\label{sec4}

In this section simulations are conducted to examine the finite-sample
performance of the proposed models. Two examples are considered. The
first example demonstrates the performance of the estimators in the
MUTARE model and the second example compares the double penalized order
selection procedure with a standard approach.

\subsection{Example 1}\label{sec4.1}

Consider the following MUTARE model with $k=2$:
\[
y_{it}=\alpha_i+\beta_0+\beta_1 y_{i, t-1} { I}[y_{i,t-1}>\tau]+\beta_2
y_{i, t-2} {I}[y_{i,t-1}>\tau, y_{i,t-2}> \tau] +\varepsilon_{it}.
\]
The coefficients of this model are fixed at $\bolds{\gamma
}_0=(\bolds{\beta}_0, 0.1, 0.5)$, where the fixed effects are
$\bolds{\beta}_0=(0,0.5,0.4)$. The random\vadjust{\goodbreak} error $\varepsilon_{it}$ is
generated from a normal distribution with mean 0 and variance $0.5$.
The sample size combinations used are $(m=30, n=10)$, $(m=40,n=15)$,
and $(m=60,n=25)$. For each combination, the simulations are conducted
based on 1000 replicates. In this example, tuning parameters are
determined by minimizing the mean squared prediction error of new
generated testing data with the same size and then fixed for all the replicates.

\begin{table}
\tablewidth=210pt
\caption{Summary of simulation results in example 1}\label{tableEX1}
\begin{tabular*}{\tablewidth}{@{\extracolsep{\fill}}lccccc@{}}
\hline
& $\bolds{\tau}$& $\bolds{\beta_0}$ & $\bolds{\beta_1}$& $\bolds{\beta_2}$
& $\bolds{\sigma^2}$ \\
\hline
\multicolumn{6}{@{}c@{}}{$m=30$, $n=10$}\\[3pt]
Mean& 0.114 & 0.038& 0.491 &0.390 &0.385\\
sd & 0.029 & 0.174 & 0.083 & 0.080 & 0.188\\
CP & &0.906& 0.859& 0.866 &\\
[5pt]
\multicolumn{6}{@{}c@{}}{$m=40$, $n=15$}\\[3pt]
Mean & 0.111& 0.035& 0.484 & 0.393 & 0.431 \\
sd & 0.028 & 0.155 & 0.058 & 0.047 & 0.140 \\
CP && 0.915& 0.868 &0.889 &\\
[5pt]
\multicolumn{6}{@{}c@{}}{$m=60$, $n=25$}\\[3pt]
Mean & 0.105 & 0.036 & 0.501 & 0.398 & 0.477 \\
sd & 0.019 & 0.123 & 0.041 & 0.032 & 0.090 \\
CP && 0.918 & 0.878 & 0.898 &\\
[5pt]
True& 0.1\hphantom{00} & 0\hphantom{000.} & 0.5\hphantom{00}
& 0.4\hphantom{00} & 0.5\hphantom{00} \\
\hline
\end{tabular*}
\end{table}

The simulation results are reported in Table~\ref{tableEX1}. For each
sample size combination, the sample means and standard deviations of
the estimates are listed. The empirical coverage probabilities of the
fixed effects, denoted by ``CP,'' are listed in the last row of each
setting. They are calculated based on the 90$\%$ confidence intervals
of the corresponding regression parameters.
As shown in the table, the sample mean of the estimates becomes closer
to the true value and the associated standard deviation becomes smaller
as the sample size increases. These results confirm the asymptotic
consistency discussed in Section~\ref{sec3}. Moreover, when the sample size
increases, the empirical coverage probabilities for the fixed effects
are closer to the nominal coverage probabilities.

\begin{figure}

\includegraphics{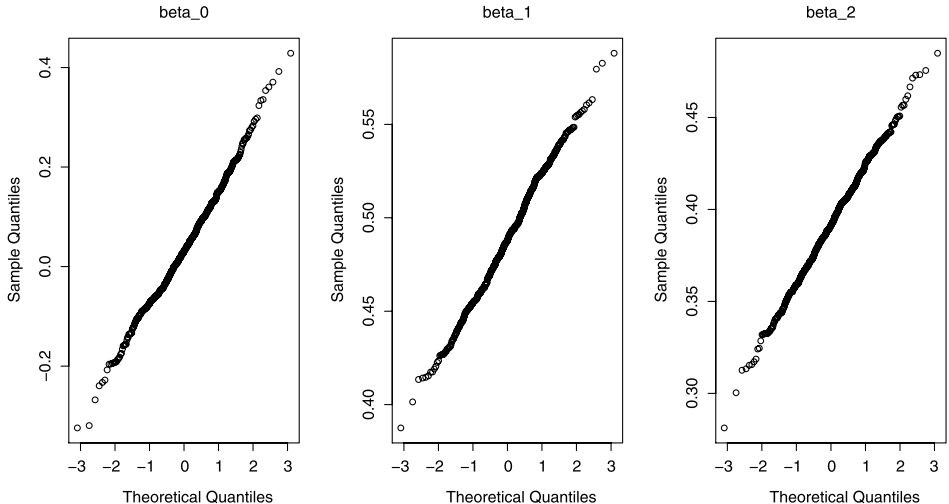}

\caption{Normal Q--Q plots in example 1.}\label{QQ1}
\end{figure}

To assess the asymptotic normality, normal Q--Q plots are reported in
Figure~\ref{QQ1}.
It is plotted based on the three estimated fixed effects, $\hat{\beta
}_1$, $\hat{\beta}_2$ and~$\hat{\beta}_3$, with the sample size
combination $m=60$ and $n=25$.
In general, the data points being close to straight lines in the Q--Q
plots confirms that the estimates are normally distributed.\eject


\subsection{Example 2}\label{sec4.2}

In this example we study the performance of the proposed order
selection procedure. Since there is no existing approach available, we
compare the double penalized approach with a naive Akaike information
criterion [AIC; \citet{Aka73}], which is suggested for order selection
in the threshold autoregressive models [\citet{Ton}], and the Bayesian
information criterion [BIC; \citet{Sch78}]. Three different models
following equation~(\ref{NLTnoRE}) are considered with parameters given
in Table~\ref{tableEX2} and sample size 200. The threshold parameters
are assumed to be 0.01 and the random errors are generated from a
normal distribution with mean 0 and variance 0.1. The tuning parameters
are determined as in example 1.

%

\begin{table}[b]
\tablewidth=150pt
\caption{Parameter values in example 2}\label{tableEX2}
\begin{tabular*}{\tablewidth}{@{\extracolsep{\fill}}lccccc@{}}
\hline
\textbf{Model} & $\bolds{\beta_1}$ & $\bolds{\beta_2}$
& $\bolds{\beta_3}$ & $\bolds{\beta_4}$
& $\bolds{\beta_5}$ \\
\hline
$1$ & $0.4 $ & $0.4 $ & $0 $\hphantom{0.}
& $0 $\hphantom{00.} &
$0 $\\
$2$ & $0.5 $ & $0.3 $ & $0.1 $& $0 $\hphantom{00.}
& $0 $\\
$3$ & $0.3 $ & $ 0.2 $ & $0.1 $& $0.05 $ &$0 $ \\
\hline
\end{tabular*}
\end{table}

Table~\ref{tableEX2result} shows the order selection performance of
AIC, BIC and the double penalized approach. The column $k_0$ indicates
the true order. For both methods, we report the percentage of times
that the estimated order equals a number of values (i.e., 1 to 5) out
of 1000 replicates. The numbers with boldface indicate the most
selected orders. For model 1, all the three methods select the right
order with their highest frequency. The double penalized approach and
BIC perform equally well in this model and both of them perform better
than AIC. For example, the double penalized approach has a 30\%
$[=(0.751-0.580)/0.580]$ higher chance to select the right order
%
\begin{table}
\caption{Simulation results in example 2}\label{tableEX2result}
\begin{tabular*}{\tablewidth}{@{\extracolsep{\fill}}lcccccc@{}}
\hline
& & \multicolumn{5}{c@{}}{\textbf{AIC}}\\[-4pt]
& & \multicolumn{5}{c@{}}{\hrulefill}\\
\textbf{Model} & $\bolds{k_0}$ & \textbf{1} & \textbf{2} & \textbf{3} & \textbf{4} & \textbf{5}\\ 
\hline
1   &   2   &   0.178            &   $\mathbf{0.580}$     &   0.193              &   0.014          &   0.020\\
2   &   3   &   0.142            &   $\mathbf{0.574}$     &   0.150              &   0.101          &   0.031\\
3   &   4   &   $\mathbf{0.522}$ &   0.325                &   0.111              &   0.042          &   0.000\\
\hline
& & \multicolumn{5}{c@{}}{\textbf{BIC}}\\[-4pt]
& & \multicolumn{5}{c@{}}{\hrulefill}\\
\textbf{Model} & $\bolds{k_0}$ & \textbf{1} & \textbf{2} & \textbf{3} & \textbf{4} & \textbf{5}\\ 
\hline
1   &   2   &   0.001            &   $\mathbf{0.749}$     &   0.152                &   0.088            &    0.001  \\
2   &   3   &   0.243            &   $\mathbf{0.536}$     &   0.151                &   0.058            &    0.012  \\
3   &   4   &   $\mathbf{0.553}$ &   0.322                &   0.110                &   0.015            &    0.000  \\
\hline
& & \multicolumn{5}{c@{}}{\textbf{Double penalized}}\\[-4pt]
& & \multicolumn{5}{c@{}}{\hrulefill}\\
\textbf{Model} & $\bolds{k_0}$ & \textbf{1} & \textbf{2} & \textbf{3} & \textbf{4} & \textbf{5}\\ 
\hline
1   &   2   &   0.103           &   $\mathbf{0.751}$      &    0.091                &   0.050               &   0.004\\
2   &   3   &   0.000           &   0.053                 &    $\mathbf{0.659}$     &   0.167               &   0.121\\
3   &   4   &   0.023           &   0.081                 &    0.248                &   $\mathbf{0.645}$    &   0.003\\
\hline
\end{tabular*}
\end{table}
compared with AIC. For models 2 and 3, both AIC and BIC tend to
underestimate the order and the double penalized approach selects the
correct order with probability higher than 65$\%$. These results
indicate that the double penalized approach outperforms the other two
methods in terms of order selection. The computational efficiency of
the double penalized approach is reasonably close to AIC and BIC in the
simulation.
The average computing times are 4.38 seconds for AIC, 4.45 seconds for
BIC and 4.92 seconds for the double penalized approach.

\section{Application in unbinding force experiments}\label{sec5}

\begin{figure}

\includegraphics{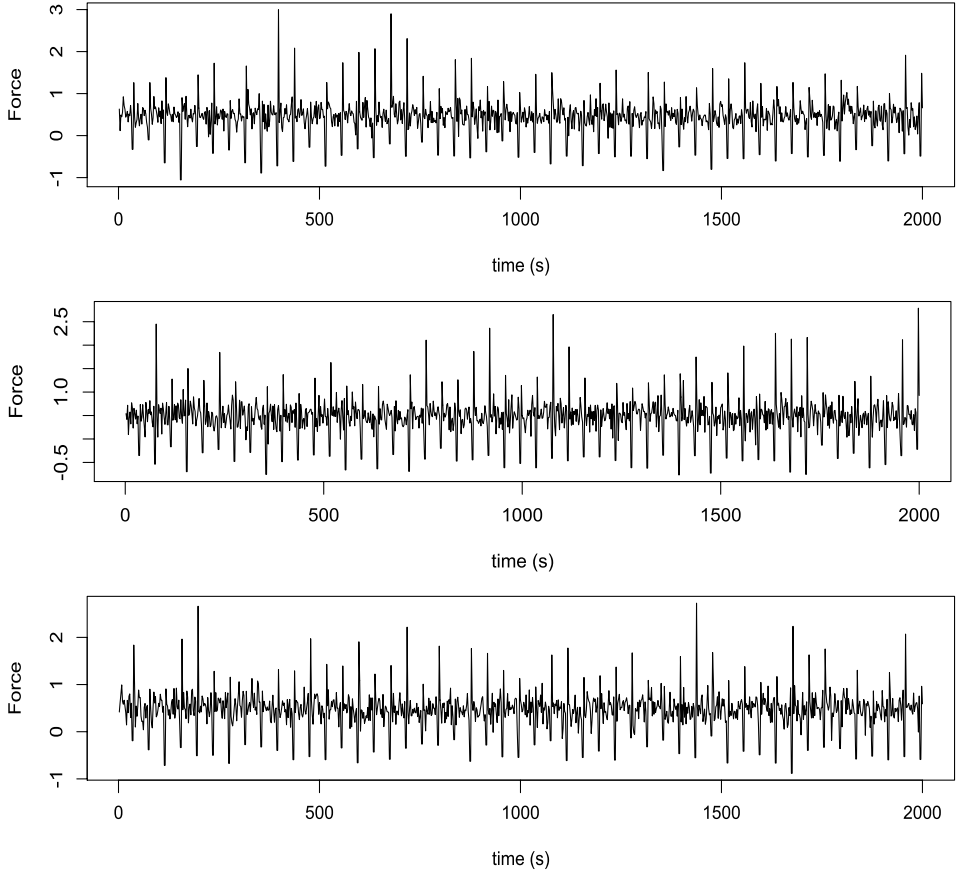}

\caption{The measurements from the repeated unbinding force
assays.}\label{realdata}
\end{figure}

In this section we revisit the repeated unbinding force experiments and
apply the proposed method to study the memory effect on
such repeated assays. There are 15 pairs of experimental subjects and
each pair includes a T-cell and a probe bead attached to a red
blood cell as described in Figure~\ref{BFP}.
For each cell adhesion cycle, a T-cell and a probe bead are brought
into contact (i.e., touch) for 4 seconds and then retracted to the
unbinding position (see Figure~\ref{Onecycle}). Such a cycle is
performed repeatedly on the same pair of experimental subjects for 50 times.
Figure~\ref{realdata} is three randomly selected samples of the
repeated unbinding forces from such experiments. For each sample, the
forces are plotted based on observations in 1000 seconds with 50
repeated adhesion cycles completed.\vadjust{\goodbreak}

The unbinding forces are collected according to the definition in
Figure~\ref{Onecycle}.
Prior knowledge [\citet{Zaretal07}, \citet{Hunetal08}]
indicates that a reasonable order of the memory in this process should
be less than 5. Therefore, we first fit the MUTARE model with $k=5$ and
then the double penalized order selection procedure is applied. The
order of the memory is identified as two and the memory effect on the
repeated unbinding force experiments can be quantified by the MUTARE
model as
\[
\hat{y}_{it}=\alpha_i+0.245 y_{i, t-1} { I}[y_{i,t-1}>\hat{\tau
}]+0.11 y_{i, t-2} {I}[y_{i,t-1}>\hat{\tau}, y_{i,t-2}> \hat{\tau}],
\]
where $i=1,\ldots, 15$, $t=1,\ldots,50$, the random effect $\alpha_i$
follows normal distribution with mean $-0.072$ and variance $0.389$.
The estimated order of the memory in this experiment is consistent with
that in \citet{Hunetal08} with a similar setting but different\vadjust{\goodbreak}
measurements. Such consistency provides important evidence of a unified
underlying kinetic mechanism in the adhesion process. The estimated
threshold, $\hat{\tau}=0.089$, indicates that an adhesion leads to a
bond only if the unbinding force is larger than $0.089 pN$. Based on
this result, the occurrence of a bond, although unobservable, can be
easily studied by measuring the corresponding unbinding forces. Since
random effects are considered, the fitted model can be used to make
inference beyond the 15 pairs of experimental subjects.

\section{Summary and concluding remarks}\label{sec6}

Despite numerous results available in modeling nonlinear time series,
their applications are limited. For example,
they are mainly constructed for a single series of observations and
focus on the case where the nonlinearity is determined
based on one variable. Furthermore, there is no order selection
procedure available with theoretical justification for such models.
Motivated by the analysis of the repeated unbinding force experiments,
a new nonlinear time series model, MUTARE, and a double penalized order
selection procedure are introduced.

The proposed model handles multiple time series by incorporating random
effects to borrow strength across different subjects. Thus, inference
and predictions can be made beyond the experimental units in the study.
Moreover, the proposed methodology provides
a new nonlinear time series model that is easy to interpret and
captures the autoregressive behavior of the observations above some
unknown threshold. The double penalized procedure can be used to
efficiently identify the order and can be easily implemented by a
global optimization algorithm using mixed integer programming. The
selection consistency and asymptotic normality of the estimators are derived.
Apart from the asymptotic results, the finite-sample performance is
examined via simulations.

As an application, the MUTARE model is illustrated by modeling the
memory effect on the repeated unbinding force assays. The fitted model
provides a better understanding of how force regulates receptor-ligand
interactions. This work is one of the first few studies considering
memory effects in the cell adhesion experiments. More studies are
needed to construct a rigorous and interpretable biological model. An
ongoing project includes theoretical development for the estimated
threshold, relaxation of the constant threshold assumption, and taking
into account important process variables, such as contact duration,
into the model.

\begin{appendix}\label{app}
\section{Assumptions}\label{appA}

\renewcommand{\theAssumption}{A\arabic{Assumption}}
\begin{Assumption}\label{AssumptionA1}
The process $y_{it}$ is stationary, ergodic and
has finite second moments.
\end{Assumption}
\begin{Assumption}\label{AssumptionA2}
The autoregressive function is discontinuous,
that is, there exists a $H^*=(1,y^*_{t-1},\ldots,y^*_{t-k})$ such that
$H^*(A_s-A_t)\neq0$ and $y_{t-1}=\cdots=y_{t-j}=\tau$, where
$A_1=(\beta_0,0,\ldots,0)',\ldots,A_{k-1}=(\beta_1,\ldots,\beta
_k)'$, $(s,t)\in(1,\ldots, k-1)$, and $j=1,\ldots,k$.
\end{Assumption}
\begin{Assumption}\label{AssumptionA3}
There exists a $M_1>0$ such that
$E [\operatorname{tr}({\mathbf W}_i^{-1} Z_iZ_i' \times{\mathbf W}_i^{-1} Z_iZ_i')]^2\leq M_1$, and
$E[\operatorname{tr} ({\mathbf W}_i^{-1} Z_iZ_i')-(
{\mathbf y}_i-g(H_i,\bolds{\beta}_0,\tau_0,\sigma^2))' {\mathbf W}_i^{-1} Z_i
Z_i'\times {\mathbf W}_i^{-1}({\mathbf y}_i-g(H_i,\bolds{\beta}_0,\tau
_0,\sigma^2) ]^2\leq M_1$ for all $i$, where ${\mathbf W}_i$ and $Z_i$ are
the matrices of the covariance and random effects for the $i$th subject.
\end{Assumption}
\begin{Assumption}\label{AssumptionA4}
$\lim\inf_{n \rightarrow\infty} \lambda
_n=\lambda>0$, where $\lambda_n$ is the smallest eigenvalue of
$-\frac{1}{n} \sum_i E[\operatorname{tr}({\mathbf W}_i^{-1} Z_iZ_i' {\mathbf W}_i^{-1}
Z_iZ_i')]$.
\end{Assumption}

Assumptions~\ref{AssumptionA1} and~\ref{AssumptionA2} are necessary for the strong consistency of the
fixed effect and threshold parameter estimators.
Assumptions~\ref{AssumptionA3} and~\ref{AssumptionA4} are used for the strong consistency of the
variance components. More discussions can be found in \citet{Nie06}.

\section{\texorpdfstring{Proof of Lemma \lowercase{\protect\ref{lemma1}}}{Proof of Lemma 1}}\label{secB}\label{appB}

The proof relies on verifying the following two claims.
\begin{Claim}\label{Claim1}
There exists a $M_2>0$ such that, for $m$
and $n$ sufficiently large, the maximum likelihood estimator of
$\bolds{\gamma}$ lies in $\Omega_2=\{\bolds{\gamma} \in\Omega
\dvtx|\beta_1-\beta_{1,0}|\leq M_2, \ldots, |\beta_k-\beta_{k,0}|\leq M_2,
|\sigma^2-\sigma^2_0|\leq M_2\}$ almost surely.
\end{Claim}
\begin{pf*}{Verification of Claim~\ref{Claim1}}
Recall $\bolds{\gamma}_0=(\bolds{\beta}_0,\tau,\sigma_0^2)$
and define $\bolds{\beta}_0=(\beta_{0,0},\ldots,\beta_{k,0})'$. To
prove Claim~\ref{Claim1}, it suffices to show that for $m$ and $n$ sufficiently
large and uniformly for $\bolds{\gamma}$ not belonging to $\Omega
_2$, we have $(mn)^{-1}(\operatorname{pl}(\bolds{\gamma})-\operatorname{pl}(\bolds{\gamma
}_0))<0$ almost surely:
%
\begin{eqnarray}\label{lemma1eq1}
\frac{\operatorname{pl}(\bolds{\gamma})-\operatorname{pl}(\bolds{\gamma}_0)}{mn}
&=&\frac
{\operatorname{pl}(\bolds{\beta},\tau,\sigma^2)-\operatorname{pl}(\bolds{\beta}_0,\tau
_0,\sigma^2)}{mn}\nonumber\\[-8pt]\\[-8pt]
&&{}+\frac{\operatorname{pl}(\bolds{\beta}_0,\tau_0,\sigma
^2)-\operatorname{pl}(\bolds{\beta}_0,\tau_0,\sigma_0^2)}{mn}.\nonumber
\end{eqnarray}
We first examine the first part on the right-hand side of (\ref
{lemma1eq1}). Assuming that the variance component is consistent along
the lines of \citet{Nie06}, the study of the first part can be transformed
into the study of $Y^*={\mathbf W}^{-1/2}(Y-Z\bolds{\alpha})$, which
is used in the derivation for both $
\operatorname{pl}(\bolds{\beta},\tau,\sigma^2)$ and $\operatorname{pl}(\bolds{\beta}_0,\tau
_0,\sigma^2)$. We have
\begin{eqnarray*}
\operatorname{pl}(\bolds{\beta},\tau,\sigma^2)-\operatorname{pl}(\bolds{\beta}_0,\tau
_0,\sigma^2)
&=& 2\log L(\bolds{\beta},\tau,\sigma^2)-2 \log L(\bolds
{\beta
}_0,\tau_0,\sigma^2)\\
&&+\sum_{j=1}^k
[ P_{\lambda_1}(|\beta_{j,0}|)-P_{\lambda_1}(|\beta_j|)
]\\
&&+\lambda_2 \Bigl[\max_j \{j\dvtx \beta_{j,0}\neq0\}-\max_j\{j\dvtx \beta_{j}\neq0\}\Bigr].
\end{eqnarray*}
First, up to an additive constant, we have
\[
2\log(\bolds{\beta}, \tau, \sigma^2)=-\sum_i \sum_{t}
\bigl(y^*_{it}-g(\bolds{\beta},\tau,\sigma^2\mid H_{it})\bigr)^2.
\]
Due to the nonlinearity, the derivation for a general MUTARE model can
be lengthy in nature.
Therefore, we illustrate the detailed derivation by a~smaller model and
consider the case where $\tau>\tau_0$. The same argument
can be easily applied and extended to the MUTARE model and the case
$\tau\leq\tau_0$ in general.%

Consider a MUTARE model with $k=2$:
%
\begin{equation}\label{MUTAREk2}\qquad
\cases{
y_{it}=\alpha_i+\beta_0+\varepsilon_{it},&\quad if
$y_{i,t-1}\leq\tau$,\cr
y_{it}=\alpha_i+\beta_0+\beta_1 y_{i,t-1}+\varepsilon_{it}, &\quad if
$y_{i,t-1}> \tau, y_{i, t-2}\leq\tau$,\cr
y_{it}=\alpha_i+\beta_0+\beta_1 y_{i,t-1}+\beta_2 y_{i,t-2}+\varepsilon
_{it}, &\quad if $y_{i,t-1}> \tau, y_{i, t-2}> \tau$,}
\end{equation}
the corresponding log likelihood function can be decomposed by
%
\begin{eqnarray}
\label{decomp}
&&2\log L(\bolds{\beta}, \tau, \sigma^2)\nonumber\\[-2pt]
&&\qquad=-\sum_{i}\sum_t (y^*_{it}-\beta_0)^2 I[y_{i,t-1}\leq\tau_0]\nonumber\\[-2pt]
&&\qquad\quad{}-\sum_i \sum_t (y^*_{it}-\beta_0)^2 I[\tau_0 < y_{i,t-1}\leq\tau,
y_{i,t-2}\leq\tau_0]\nonumber\\[-2pt]
&&\qquad\quad{}-\sum_i \sum_t (y^*_{it}-\beta_0)^2 I[\tau_0 < y_{i,t-1}\leq\tau,
y_{i,t-2}> \tau_0]\nonumber\\[-9.5pt]\\[-9.5pt]
&&\qquad\quad{}-\sum_i \sum_t (y^*_{it}-\beta_0-\beta_1 y^*_{i,t-1})^2 I[y_{i,t-1}>
\tau, y_{i,t-2}\leq\tau_0]\nonumber\\[-2pt]
&&\qquad\quad{}-\sum_i \sum_t (y^*_{it}-\beta_0-\beta_1 y^*_{i,t-1})^2 I[y_{i,t-1}>
\tau, \tau_0 < y_{i,t-2}\leq\tau]\nonumber\\[-2pt]
&&\qquad\quad{}-\sum_i \sum_t (y^*_{it}-\beta_0-\beta_1 y^*_{i,t-1}-\beta_2
y^*_{i,t-2})^2 I[y_{i,t-1}> \tau,y_{i,t-2}> \tau]\nonumber\\[-2pt]
&&\qquad= R_1 (\bolds{\beta},\tau,\sigma^2)+\cdots+ R_6 (\bolds{\beta
},\tau,\sigma^2).\nonumber
\end{eqnarray}

Defining $A_1=(\beta_{0,0},0,0)'$, $A_2=(\beta_{0,0},\beta_{1,0},0)'$,
$A_3=(\beta_{0,0},\beta_{1,0},\beta_{2,0})'$, $B_1=(\beta_0,0,0)'$,
$B_2=(\beta_0,\beta_1,0)'$, and $B_3=(\beta_0,\beta_1,\beta_2)'$, we have
\begin{eqnarray*}
&&R_4(\bolds{\beta},\tau,\sigma^2)-R_4(\bolds{\beta}_0,\tau
_0,\sigma^2)\\[-2pt]
&&\qquad=\sum_i\sum_t [ -(y^*_{it}-\beta_0-\beta_1
y^*_{i,t-1})^2+(y^*_{it}-\beta_{0,0}-\beta_{1,0} y^*_{i,t-1} )^2]\\[-2pt]
&&\qquad\quad\hspace*{27pt}{}\times I(y_{i,t-1}> \tau, y_{i,t-2}\leq\tau_0)\\[-2pt]
&&\qquad= \sum_i\sum_t [ -(y^*_{it}-
H^*_{i,t-1}B_2)^2+(y^*_{it}-H^*_{i,t-1}A_2 )^2]\\[-2pt]
&&\qquad\quad\hspace*{27pt}{}\times  I(y_{i,t-1}> \tau,
y_{i,t-2}\leq\tau_0)\\[-2pt]
&&\qquad=2|B_2-A_2|\sum_i\sum_t H^*_{i,t-1} \frac{(B_2-A_2)}{|B_2-A_2|}
(y^*_{it}-H^*_{i,t-1}A_2)\\[-2pt]
&&\qquad\quad\hspace*{80.1pt}{}\times I(y_{i,t-1}> \tau, y_{i,t-2}\leq\tau_0)\\[-2pt]
&&\qquad\quad{}-|B_2-A_2|^2 \sum_i \sum_t \biggl(H^*_{i,t-1}\frac
{(B_2-A_2)}{|B_2-A_2|}\biggr)^2I(y_{i,t-1}> \tau, y_{i,t-2}\leq\tau_0).
\end{eqnarray*}
Therefore, based on the uniform law of large numbers [\citet{Pol84},
page~8], we have
\begin{eqnarray*}
&&\frac{1}{mn}\bigl(\operatorname{pl}(\bolds{\beta},\tau,\sigma^2)-\operatorname{pl}(\bolds{\beta
}_0,\tau_0,\sigma^2)\bigr)\\[-2pt]
&&\qquad\leq 2(|B_1-A_1|+|B_1-A_2|+|B_1-A_3|\\[-2pt]
&&\hspace*{42.6pt}{}+|B_2-A_2|+|B_2-A_3|+|B_3-A_3|)
\varepsilon\\[-2pt]
&&\qquad\quad{}
-(|B_1-A_1|^2+|B_1-A_2|^2+|B_1-A_3|^2\\[-2pt]
&&\qquad\quad\hspace*{17.2pt}{}+|B_2-A_2|^2+|B_2-A_3|^2+|B_3-A_3|^2)(K-\varepsilon
)\\[-2pt]
&&\qquad\quad{}+(mn)^{-1}\Biggl\{\sum_{j=1}^k
[ P_{\lambda_1}(|\beta_{j,0}|)-P_{\lambda_1}(|\beta_j|)
]\\[-2pt]
&&\hspace*{52pt}\qquad\quad{}
+\lambda_2 \Bigl[\max_j \{j\dvtx \beta_{j,0}\neq0\}-\max_j\{j\dvtx\beta_{j}\neq0\}
\Bigr]\Biggr\}\\[-2pt]
&&\qquad=2 \varepsilon\Delta_1-\Delta_2 (K-\varepsilon)+\Delta_3,
\end{eqnarray*}
where
\[
K=\inf_{\beta} \min_{i \leq j} E \biggl( \biggl(H_{i,t-1}\frac
{(B_i-A_j)}{|B_i-A_j|}\biggr)^2I_{ij}\biggr)
\]
and $I_{ij}$ is the corresponding indicator function as listed in (\ref
{decomp}). Note that the uniform law of large numbers in Pollard
[(\citeyear{Pol84}), page 8] assumes that the data are independent and identically
distributed. This assumption is relaxed to a stationary ergodic process
by \citet{SamCha11}. Therefore, the uniform law of large numbers
can be applied here.
Based on the Cauchy--Schwarz inequality, we have $\Delta_1 \leq\sqrt{6
\Delta_2} \leq6\Delta_2$ for sufficiently large $M_2$. For
sufficiently large $m$ and~$n$, $\Delta_3<\varepsilon\Delta_2$. Thus,
by selecting $\varepsilon<K/14$, it follows that
$(mn)^{-1}(l(\bolds
{\beta},\tau,\sigma^2)-l(\bolds{\beta}_0,\tau_0,\sigma^2))<0$.

For the second term on the right-hand side of~(\ref{lemma1eq1}), under
Assumptions~\ref{AssumptionA3} and~\ref{AssumptionA4}, the maximum
likelihood estimator of the variance component almost surely converges
based on the results in \citet{Nie06}. Therefore,\vspace*{1pt} we
have $(mn)^{-1}(l(\bolds{\beta}_0,\tau
_0,\sigma^2)-l(\bolds{\beta}_0,\tau_0,\sigma_0^2))<0$ and Claim
\ref{Claim1} follows. \noqed\end{pf*}
\begin{Claim}\label{Claim2}
There exists a $M_3>0$ such that, for $m$
and $n$ sufficiently large, the maximum likelihood estimator of
$\bolds{\gamma}$ lies in $\Omega_3=\{\bolds{\gamma} \in\Omega
_2\dvtx |\tau-\tau_{0}|\leq M_3 \}$ almost surely.\vadjust{\goodbreak}
\end{Claim}
\begin{pf*}{Verification of Claim~\ref{Claim2}}
Similar to Claim~\ref{Claim1}, it suffices to show that, for $m$ and $n$
sufficiently large, $(mn)^{-1}(\operatorname{pl}(\bolds{\gamma})-\operatorname{pl}(\bolds
{\gamma}_0))<0$ for $\bolds{\gamma}$ not belonging to $\Omega_3$.
We apply the same decomposition as in Lemma~\ref{lemma1} and focus on the first
part on the right-hand side of~(\ref{lemma1eq1}). Applying the uniform
law of large numbers and the same transformation as described in Claim
\ref{Claim1}, for $m$ and $n$ sufficiently large, it holds that
\begin{eqnarray*}
&&\frac{\operatorname{pl}(\bolds{\beta},\tau,\sigma^2)-\operatorname{pl}(\bolds{\beta}_0,\tau
_0,\sigma^2)}{mn}\\[-0.5pt]
&&\qquad= {mn}^{-1}\Biggl\{2\log L(\bolds{\beta},\tau,\sigma^2)\\[-0.5pt]
&&\hspace*{65pt}{}-2 \log
L(\bolds{\beta}_0,\tau_0,\sigma^2)\sum_{j=1}^k
[ P_{\lambda_1}(|\beta_{j,0}|)-P_{\lambda_1}(|\beta_j|)
]\\[-0.5pt]
&&\hspace*{65pt}\hspace*{31pt}{} +\lambda_2 \Bigl[\max_j \{j\dvtx \beta_{j,0}\neq0\}-\max_j\{j\dvtx \beta_{j}\neq0\}
\Bigr]\Biggr\}\\[-0.5pt]
&&\qquad\leq E\bigl\{\bigl(-(y^*_{it}-H^*_{i,t-1}B_1)^2+(y_{it}-H_{i,t-1}A_1)^2\bigr)
I[y_{i,t-1}\leq\tau_0]\bigr\}\\[-0.5pt]
&&\qquad\quad{}+E\bigl\{\bigl(-(y^*_{it}-H^*_{i,t-1}B_1)^2+(y^*_{it}-H^*_{i,t-1}A_2)^2\bigr)\\
&&\qquad\quad\hspace*{68pt}{}\times I[\tau
_0 < y_{i,t-1}\leq\tau, y_{i,t-2}\leq\tau_0]\bigr\}\\[-0.5pt]
&&\qquad\quad{}+E\bigl\{\bigl(-(y^*_{it}-H^*_{i,t-1}B_1)^2+(y^*_{it}-H^*_{i,t-1}A_3)^2\bigr)\\
&&\qquad\quad\hspace*{68pt}{}\times I[\tau
_0 < y_{i,t-1}\leq\tau, y_{i,t-2}> \tau_0]\bigr\}\\[-0.5pt]
&&\qquad\quad{}+E\bigl\{\bigl(-(y^*_{it}-H^*_{i,t-1}B_2)^2+(y^*_{it}-H^*_{i,t-1}A_2)^2\bigr)\\
&&\qquad\quad\hspace*{89pt}{}\times
I[y_{i,t-1}> \tau, y_{i,t-2}\leq\tau_0]\bigr\}\\[-0.5pt]
&&\qquad\quad{}+E\bigl\{\bigl(-(y^*_{it}-H^*_{i,t-1}B_2)^2+(y^*_{it}-H^*_{i,t-1}A_3)^2\bigr)\\
&&\qquad\quad\hspace*{70pt}{}\times
I[y_{i,t-1}> \tau, \tau_0 < y_{i,t-2}\leq\tau]\bigr\}\\[-0.5pt]
&&\qquad\quad{}+E\bigl\{\bigl(-(y^*_{it}-H^*_{i,t-1}B_3)^2+(y^*_{it}-H^*_{i,t-1}A_3)^2\bigr)\\
&&\qquad\quad\hspace*{92pt}{}\times I[y_{i,t-1}> \tau,y_{i,t-2}> \tau]\bigr\}+\varepsilon.
\end{eqnarray*}
Considering the situation where $\tau>\tau_0$, we have
\[
\frac{l(\bolds{\beta},\tau,\sigma^2)-l(\bolds{\beta}_0,\tau
_0,\sigma^2)}{mn}\leq J+\varepsilon,
\]
where
\begin{eqnarray*}
J&=&E\bigl\{\bigl(-(y^*_{it}-H^*_{i,t-1}B_1)^2+(y^*_{it}-H^*_{i,t-1}A_1)^2\bigr)
I[y_{i,t-1}\leq\tau_0]\bigr\}\\[-0.5pt]
&&{}+E\bigl\{\bigl(-(y^*_{it}-H^*_{i,t-1}B_1)^2+(y^*_{it}-H^*_{i,t-1}A_2)^2\bigr)\\
&&\qquad\quad\hspace*{34pt}{}\times I[\tau
_0 < y_{i,t-1}\leq\tau, y_{i,t-2}\leq\tau_0]\bigr\}\\[-0.5pt]
&&{}+E\bigl\{\bigl(-(y^*_{it}-H^*_{i,t-1}B_1)^2+(y^*_{it}-H^*_{i,t-1}A_3)^2\bigr)\\
&&\qquad\quad\hspace*{34pt}{}\times I[\tau
_0 < y_{i,t-1}\leq\tau, y_{i,t-2}> \tau_0]\bigr\}.
\end{eqnarray*}
When $\tau=\infty$, the model becomes a linear mixed model; therefore,
by the dominated convergence theorem and
a similar argument in \citet{SamCha11}, it holds almost surely
that, for $m$ and $n$ sufficiently large and for any $M_3>0$,
$(mn)^{-1}(l(\bolds{\beta},\tau,\sigma^2)-l(\bolds{\beta
}_0,\tau_0,\sigma^2))<0$ for $\tau\geq\tau_0+M_3$.
Similar derivation can be applied to the case $\tau<\tau_0$, thus the
detail is omitted.

Following the same argument for Claim~\ref{Claim1}, the second part on the
right-hand side of~(\ref{lemma1eq1}) is smaller than 0 with Assumptions
\ref{AssumptionA3} and~\ref{AssumptionA4}. Therefore, Lemma~\ref{lemma1} holds.
\noqed\end{pf*}

\section{\texorpdfstring{Proof of Theorem \lowercase{\protect\ref{theorem1}}}{Proof of Theorem 1}}\label{secC}\label{appC}

Without loss of generality, the parameter space
can be restricted to $\Omega_{\delta}=\{\bolds{\gamma}
\in\Omega\dvtx
|\bolds{\beta}-\bolds{\beta}_0|<\delta, |\sigma^2-\sigma
_0^2|<\delta, |\tau-\tau_0|<\delta\}$ according to Lemma~\ref{lemma1}. To simplify
the notation, we assume that $\tau_0=0$.
Because the derivation for a general model is lengthy, we consider the
same model in Lemma~\ref{lemma1}, the MUTARE model with $k=2$ in~(\ref{MUTAREk2}),
and assuming $\tau>0$, we have
\begin{eqnarray*}
&&
\operatorname{pl}(\bolds{\beta},\tau,\sigma^2)-\operatorname{pl}(\bolds{\beta},0,\sigma
^2)\\
&&\qquad=2\log L(\bolds{\beta},\tau,\sigma^2)-2\log L(\bolds{\beta
},0,\sigma^2)\\
&&\qquad=-\sum_i\sum_t \{
[(y_{it}^*-H_{i,t-1}^*B_1)^2-(y_{it}^*-H_{i,t-1}^*B_2)^2 ] Q_1 \\
&&\qquad\quad\hspace*{43pt}{}+[(y_{it}^*-H_{i,t-1}^*B_1)^2-(y_{it}^*-H_{i,t-1}^*B_3)^2] Q_2 \\
&&\qquad\quad\hspace*{44pt}{}+[(y_{it}^*-H_{i,t-1}^*B_2)^2-(y_{it}^*-H_{i,t-1}^*B_3)^2] Q_3\}
\\
&&\qquad\leq - \sum_i \sum_t \bigl\{\bigl[2H^*_{i,t-1}(B_2-B_1)\varepsilon
_{it}+\bigl(H^*_{i,t-1}(A_2-B_1)\bigr)^2\\
&&\hspace*{172.5pt}{}-\bigl(H^*_{i,t-1}(A_2-B_2)\bigr)^2\bigr] Q_1\\
&&\qquad\quad\hspace*{43pt}{} +\bigl[2H^*_{it-1}(B_3-B_1)\varepsilon
_{it}+\bigl(H^*_{i,t-1}(A_3-B_1)\bigr)^2\\
&&\hspace*{183pt}{}-\bigl(H^*_{i,t-1}(A_3-B_3)\bigr)^2\bigr] Q_2\\
&&\qquad\quad\hspace*{43pt}{} +\bigl[2H^*_{it-1}(B_3-B_2)\varepsilon
_{it}+\bigl(H^*_{i,t-1}(A_3-B_2)\bigr)^2\\
&&\hspace*{185pt}{}-\bigl(H^*_{i,t-1}(A_3-B_3)\bigr)^2\bigr] Q_3\bigr\},
\end{eqnarray*}
where $Q_1=I(0<y_{i,t-1}\leq\tau, y_{i,t-2}\leq0)$,
$Q_2=I(0<y_{i,t-1}\leq\tau, y_{i,t-2}>0)$, $Q_3=I(\tau<y_{i,t-1},
0<y_{i,t-2}\leq\tau)$.
If $\delta$ is sufficiently small, based on Assumption~\ref{AssumptionA2}, we
have\vspace*{1pt} $\sum
_i \sum_j [ (H^*_{i,t-1}(A_s-B_j))^2-(H^*_{i,t-1}(A_s-B_s))^2] Q_k \geq
0$, for $k=1,2,3$ and $s>j$. Therefore, by the same argument in
Proposition 1 of \citet{Cha93}, it holds that
for all $\varepsilon>0$, there exists a $T$ such that with probability
greater than $1-\varepsilon$,
$\bolds{\gamma}\in\Omega_{\delta}$, $\tau>T/N$, implies
$l(\bolds{\beta},\tau,\sigma^2)-l(\bolds{\beta},0,\sigma^2)<0$.
Similar derivation can be extended to the case where $\tau<-T/N$.
Hence, Theorem~\ref{theorem1} holds.

\section{\texorpdfstring{Proof of Theorem \lowercase{\protect\ref{theorem2}}}{Proof of Theorem 2}}\label{secD}\label{appD}

We first prove the asymptotic normality. Based on the adaptive lasso penalty,
\[
\hat{{\mathbf u}}=\mathop{\arg\min}_{{\mathbf u}} npl({\mathbf u}),
\]
where
$npl({\mathbf u})=-2\log L(\bolds{\beta}+{\mathbf u},\sigma^2,\tau)+\lambda
_1\sum_{j=1}^k \nu_j(|\beta_j+u_j|)+\lambda_2 \max_j\{j\dvtx \beta_j+u_j\neq
0\}$.
By the Taylor expansion, we have
\begin{eqnarray*}
npl({\mathbf u})&=&npl({\mathbf0})-{\mathbf u}'\tilde{H}'{\mathbf W}^{-1}({\sigma
})\bigl(Y-g(H,\bolds{\beta},\tau,\sigma^2)\bigr)\\
&&{}+\frac{1}{2}{\sqrt{N}\mathbf u}'
\biggl( \frac{\tilde{H} {\mathbf W}^{-1}({\sigma}) \tilde{H}'}{N} \biggr) {\sqrt{N}\mathbf
u}\\
&&{}+\lambda_1\sum_{j=1}^k \nu_j(|\beta_j+u_j|-|\beta_j|)\\
&&{}+\lambda_2 \Bigl( \max
_j\{j\dvtx \beta_j+u_j\neq0\} -\max_j\{j\dvtx \beta_j\neq0\} \Bigr).
\end{eqnarray*}
The last term on the right-hand side equals 0 if $u_j=0$ and $\beta
_j=0$, combining with the fact that [\citet{Zou06}]
%
\begin{equation}
\lambda_1 \nu_j (|\beta_j+u_j|-|\beta_j|)
\rightarrow_{\mathcal{P}} \cases{
0, &\quad if $\beta_j \neq0$,\cr
0, &\quad if $\beta_j = 0 \mbox{ and } u_j=0$,\cr
\infty, &\quad if $\beta_j =0 \mbox{ and } u_j \neq0$,}
\end{equation}
we have for every ${u}$
\begin{eqnarray*}
&&
npl( {\mathbf u})-npl({\mathbf0})\\
&&\qquad\rightarrow_{D} \cases{
\displaystyle - {\mathbf u}'_{(1)}\tilde{H}(1)'{\mathbf W}^{-1}({\sigma})\bigl(Y-g(H,\bolds
{\beta},\tau,\sigma^2)\bigr)\vspace*{2pt}\cr
\qquad{} +\displaystyle \frac{(\sqrt{N} {\mathbf u}_{(1)})'\Lambda
_{11}(\sqrt{N} {\mathbf u}_{(1)})}{2}, \qquad \mbox{if ${u}_{(2)}={ 0}$},
\vspace*{2pt}\cr
\infty, \qquad\hspace*{134.2pt} \mbox{otherwise.}}
\end{eqnarray*}
By the same argument of Theorem 2 in \citet{Zou06}, the asymptotic
normality holds by the martingale central limit theorem [\citet{HalHey80}].

For consistency, it suffices to show that $P(\hat{
{\beta}}_{(2)}\neq{ 0})\rightarrow0$. Using the Karush--Kuhn--Tucker
(KKT) optimality
conditions, it follows that
\[
2\tilde{H}(1)'{\mathbf W}^{-1}({\sigma})\bigl(Y-g(H,\hat{\bolds{\beta}},\tau
,\sigma^2)\bigr)=\lambda_1
{\nu}_{(1)},
\]
where $ {\nu}_{(1)}$ are the weights corresponding to the first $q$
variables. Note that $\lambda\frac{
{\nu}_{(1)}}{\sqrt{N}}\rightarrow_{\mathcal{P}}\infty$ [Theorem 2,
\citet{Zou06}] and
$
2\frac{\tilde{H}(1)'{\mathbf W}^{-1}({\sigma})(Y-g(H,\hat{\bolds{\beta
}},\tau,\sigma^2))}{\sqrt{N}}
$ is asymptotically normal. Therefore,
\[
P\bigl(\hat{ {\beta}}_{(2)}\neq{ 0}\bigr)\leq P\bigl(2\tilde{H}(1)'{\mathbf
W}^{-1}({\sigma})\bigl(Y-g(H,\hat{\bolds{\beta}},\tau,\sigma^2)\bigr)=\lambda_1
{\nu}_{(1)}\bigr)\rightarrow0,
\]
and Theorem~\ref{theorem2} holds.
\end{appendix}

\section*{Acknowledgments}

The author is grateful to the Editor, Associate Editor and two
referees for their helpful comments and suggestions, and the author
would like to thank C. F. Jeff Wu and C. Zhu for helpful discussions.



\printaddresses

\end{document}